\documentclass[aps,prl,twocolumn,superscriptaddress,groupedaddress]{revtex4}
\usepackage{graphicx}
\usepackage[english]{babel}
\usepackage{bm}       
\usepackage{amssymb}  
\usepackage{amsmath}
\usepackage{mathrsfs}
\usepackage{hyperref}
\usepackage{xcolor}

\newcommand{\red}[1]{#1}

\begin{document}

\title{Emergence of oscillons in kink-impurity interactions}

\author{Mariya Lizunova}
\affiliation{Institute for Theoretical Physics, Utrecht University, Princetonplein 5, 3584 CC Utrecht, The Netherlands}
\affiliation{Institute for Theoretical Physics Amsterdam, University of Amsterdam, Science Park 904, 1098 XH Amsterdam, The Netherlands}
\author{ Jasper Kager}
\affiliation{Institute for Theoretical Physics Amsterdam, University of Amsterdam, Science Park 904, 1098 XH Amsterdam, The Netherlands}
\author{Stan de Lange}
\affiliation{Institute for Theoretical Physics Amsterdam, University of Amsterdam, Science Park 904, 1098 XH Amsterdam, The Netherlands}
\author{ Jasper van Wezel}
\affiliation{Institute for Theoretical Physics Amsterdam, University of Amsterdam, Science Park 904, 1098 XH Amsterdam, The Netherlands}


\begin{abstract}
  The $(1+1)$-dimensional classical $\varphi^4$ theory contains stable, topological excitations in the form of solitary waves or kinks, as well as \red{a non-topological one}, such as the oscillon. Both are used in effective descriptions of excitations throughout myriad fields of physics. The oscillon is well-known to be a coherent, particle-like \red{structure} when introduced as an Ansatz in the $\varphi^4$ theory. Here, we show that oscillons also arise naturally in the dynamics of the theory, in particular as the result of kink-antikink collisions in the presence of an impurity. We show that in addition to the scattering of kinks and the formation of a breather, both bound oscillon pairs and propagating oscillons may emerge from the collision. We discuss their resonances and critical velocity as a function of impurity strength and highlight the role played by the impurity in the scattering process.
\end{abstract}

\maketitle

\section{Introduction}
The classical $\varphi^4$ theory emerges as an effective description of processes throughout fields of physics, going well beyond its original introduction as a phenomenological theory of second-order phase transitions~\cite{gltheory}. In particular, the solitary wave solutions of the theory, known as kinks, feature in the effective description of domain walls in molecules, solid materials, and even cosmology~\cite{bishop, bishop2,mele, friedland}, as well as in various toy models in nuclear physics~\cite{campbell1, wick, boguta} and biophysics~\cite{protein1,protein2}. The kink is a stable, particle-like excitation that is protected from decay by its topological charge. In addition, the $\varphi^4$ theory contains well-known non-topological excitations, in the form of  breathers and oscillons. Both are quasi-long-lived oscillating \red{structures} with applications as effective descriptions of objects throughout physics.

\red{Interactions between kinks have recently been at the centre of interest in various models and circumstances \cite{referee1,referee2,referee10,referee11,referee23,referee3,referee4,referee5,referee12,referee6,referee7,referee8,referee9}.} Unlike solitons in the integrable sine-Gordon model~\cite{aek}, kinks and antikinks \red{in the $\varphi^4$ model} cannot just pass through each other~\cite{CampbellBook}, but instead interact and undergo dynamic processes including scattering, the formation of bound states, and resonances~\cite{ablowitz, anninos, goodman2, lizunova02,lizunova03}. \red{Breathers (also called bions in $(1+1)$-dimensional $\varphi^4$)} are formed naturally as post-collision bound states \red{of the two kinks} in simulations of colliding kink-antikink pairs. \red{Another quasi-long-lived oscillating structure is the oscillon. It can exist as a localized oscillating structure on its own, without a kink-antikink collision, and the oscillations persist for a long time~\cite{referee13}. In contrast to the breathers,} oscillons usually do not emerge naturally from kink-antikink collisions \red{in the $\varphi^4$ model}. Usually they are explicitly introduced as an Ansatz for the field configuration~\cite{oscillon4,oscillon5}. Besides, they have only been identified in the dynamics of a sinh-deformed $(1+1)$-dimensional $\varphi^4$ model~\cite{oscillon8}, after applying a model deformation~\cite{oscillon9,oscillon10}, \red{and within a parametrized Dikand\'e-Kofan\'e (DW) potential with a real parameter that controls its shape profile and for some values reduces it to the standard $\varphi^4$ potential~\cite{referee14}.}

In this work, we show that oscillons do emerge naturally as a product of kink-antikink collision dynamics in the  $(1+1)$-dimensional $\varphi^4$ model, if the collision occurs in the presence of an impurity or defect. Such impurities are unavoidable in the atomic lattices of condensed matter and quantum chemistry settings, while in particle physics and cosmology, variations in the potential or background metric may be modeled by the introduction of impurities~\cite{kivshar,konotop,javidan,javidan01,javidan02,javidan03,askari}. Although the interaction of a single kink with an impurity is well-studied, interactions between kink-antikink pairs and impurities have been considered only in the context of the integrable sine-Gordon model \cite{ekomasov,malomed3,malomed4}. Here, we show that the kink-impurity-antikink dynamics not only give rise to excitation of the impurity mode and capture, resonance, and reflection of the kink-antikink pair, but that the presence of an impurity also catalyzes the formation of pairs of oscillons at the impurity site. These are observed as quasi-long-lived \red{structures} both in the form of bound oscillon-oscillon pairs and in configurations where the two oscillons independently propagate away from the impurity location.

\section{Model}
We consider a classical real scalar field $\varphi(t,x)$ in $(1+1)$-dimensional spacetime. The Lagrangian in the presence of a point-like impurity located at $x=0$ yields the equation of motion:
\begin{align}
\varphi_{tt}-\varphi_{xx}+(\varphi^3-\varphi)(1-\epsilon\gamma (x))=0. \notag
\end{align}
Here, $|\epsilon|$ represents the strength of the impurity potential $\gamma(x)$ centered at $x=0$. Note that the field dynamics in the pristine case $\epsilon=0$ is well-studied~\cite{campbell}, and already gives rise to a rich phenomenology containing breathers and resonances. Below, we discuss the field dynamics in the presence of a weak impurity, i.e. for $|\epsilon|<1$. To establish a connection to analytic results for the excitation modes of isolated Dirac-delta impurities~\cite{kivshar}, we consider an impurity profile with a very narrow Gaussian shape:
\begin{align}
\gamma (x) = \frac{1}{\sigma\sqrt{2\pi}}\text{exp}\left[-\left(\frac{x}{\sigma\sqrt{2}}\right)^2\right]. \notag
\end{align}
We choose the width of the Gaussian to be $\sigma\simeq 0.016$, following Ref.~\cite{kivshar}, to ensure that it is smaller than the lattice spacing in our numerical calculation, in accordance with the Dirac-delta limit.

\begin{figure}[tb]
\includegraphics[width=\columnwidth]{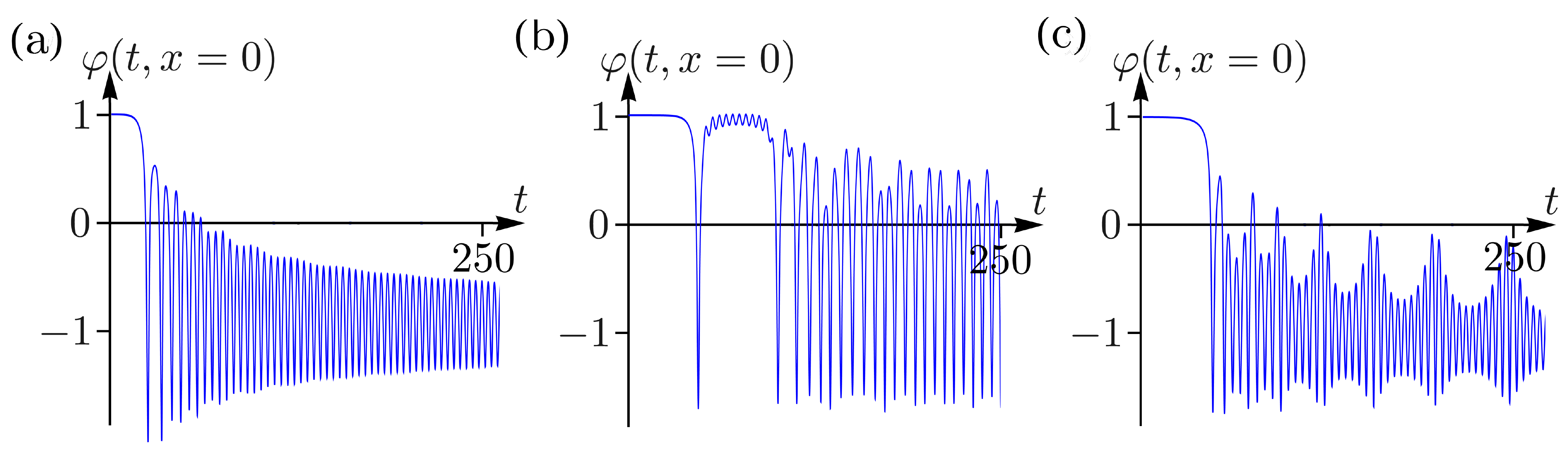}
\caption{Different types of bound states for subcritical initial velocities in kink-impurity-antikink collisions. The dynamics transitions from (a) a fast decay into the vacuum state (shown at $\epsilon=0.4$ and $v_{\text{in}}=0.2$) through (b) a quasi-long-lived breather ($\epsilon=0.15$ and $v_{\text{in}}=0.15$) to (c) an oscillon-oscillon bound state ($\epsilon=-0.2$ and $v_{\text{in}}=0.1$).}
\label{fig:vcr_epsilon_transition}
\end{figure}

We study the evolution of a kink-antikink pair, starting from the initial condition with a kink and antikink moving towards each other:
\begin{align}
\varphi(t,x)=-1 &+ \tanh\left(\frac{x+a-v_{\text{in}}\,t}{\sqrt{2(1-v_{\text{in}}^2)}}\right) \notag \\
&-\tanh\left(\frac{x-a+v_{\text{in}}\,t}{\sqrt{2(1-v_{\text{in}}^2)}}\right).\notag
\end{align}
Here, $\pm a$ are centers of localization for the kink and antikink at $t=0$, and $\mp v_{\text{in}}$ are their initial velocities. We use natural units, so that $0<v_{\text{in}}<1$. The equation of motion for the field can be solved numerically using a method of finite differences for open boundary conditions~\cite{Courant}, with the continuity equation verified at each time step.

\section{Attractive impurity}
We first consider the case of the kink and antikink colliding at the site of an attractive impurity, with $\epsilon>0$. If the initial velocity $|v_{\text{in}}|$ is smaller than a critical velocity $v_{\text{cr}}$, a breather is formed after the collision, similar to what occurs in kink-antikink scattering without an impurity present. \red{(Note that $v_{\text{cr}}$ depends on the parameter $\epsilon$, as discussed below.)} However, in the presence of the attractive impurity, the breather is no longer a \red{quasi-long-lived} and the field configuration quickly decays to the vacuum solution $\varphi(x)=-1$ (as shown in Figs.~\ref{fig:vcr_epsilon_transition}a and~\ref{fig:vcr_epsilon_transition}b). Resonances occur for special values of $v_{\text{in}}<v_{\text{cr}}$, at which the kink and the antikink leave each other after a finite number of impacts. Note that we observe resonances for various combinations of impurity strength and initial velocity, such as $\epsilon=0.05$ and $v_{\text{in}}=0.210$, or $\epsilon=0.15$ and $v_{\text{in}}=0.153$.

\begin{figure}[tb]
\begin{center}
\includegraphics[width=0.8\columnwidth]{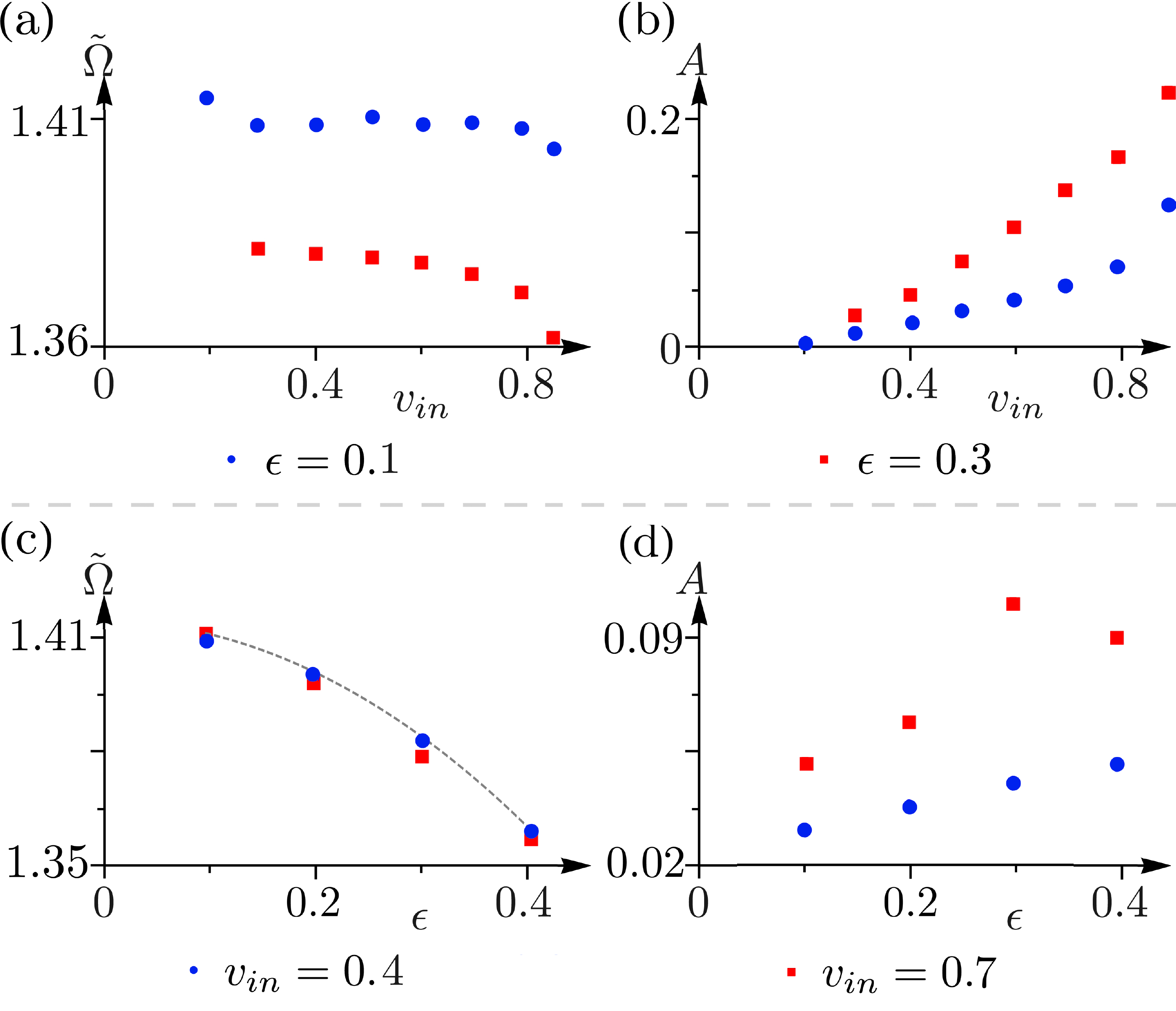}
\end{center}
\caption{(a) The impurity mode frequency $\tilde\Omega$ and (b) its amplitude $A$, as a function of the initial kink and antikink velocity $v_{\text{in}}$, for fixed values of impurity strength. \red{In (a) and (b) the blue dots and red squares correspond to $\epsilon=0.1$ and $\epsilon=0.3$ respectively.} (c) The frequency and (d) amplitude as a function of the impurity strength $\epsilon$, keeping the initial velocity constant. \red{In (c) and (d) the blue dots and red squares correspond to $v_{\text{in}}=0.4$ and $v_{\text{in}}=0.7$ respectively.} }
\label{fig:1_dependance}
\end{figure}

\begin{figure*}[tb]
\center{\includegraphics[width=0.9\linewidth]{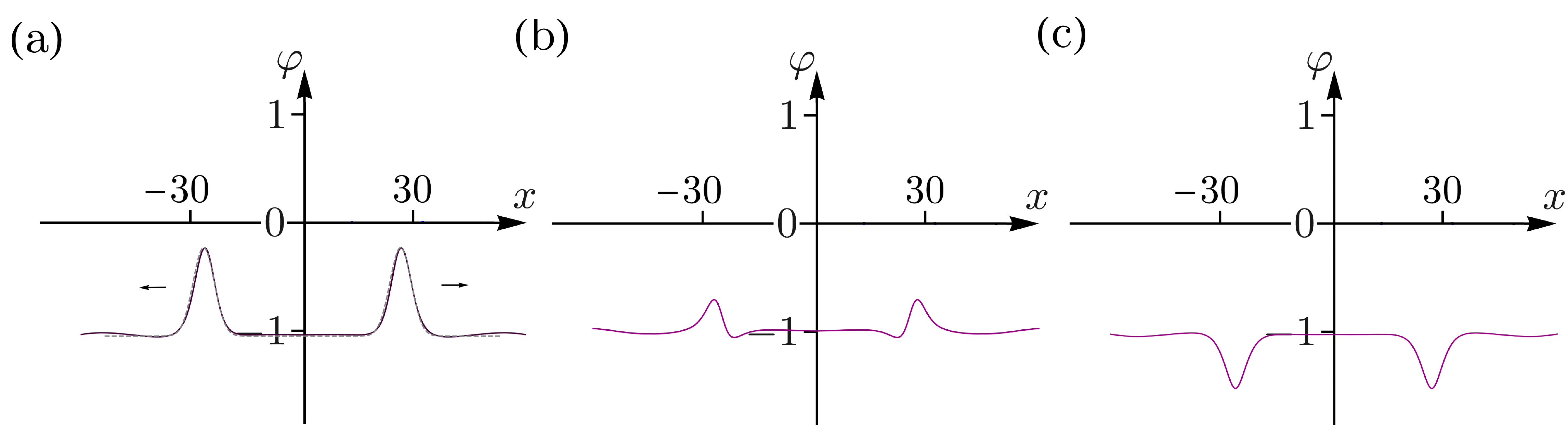}}
\caption{Moving pair of oscillons observed for $\epsilon=-0.2$ and $v_{\text{in}}=0.3$. The arrows in panel (a) represent the direction of motion of the oscillons, and the dashed line is a fit of the profile to the Ansatz of Eq.~\eqref{eq:oscillon}. Panels (b) and (c) indicate how the profile evolves over time. \red{All panels show the quality behavior at a representative moment after the kink-impurity-antikink collision.}}
\label{fig:oscillon_fig_type2}
\end{figure*}

When the kink and the antikink collide at the impurity site with initial velocities greater than $v_{\text{cr}}$, they leave the impurity location after a single collision and propagate to spatial infinity. They leave behind an oscillating mode, localized at the site of the impurity, $x=0$. Similar impurity modes have been observed in numerical simulations of kink-impurity interactions~\cite{kivshar,ourownpreprint}, and found analytically for a Dirac-delta impurity~\cite{kivshar}, where its time-dependent profile $\delta\varphi(t,x)$ was predicted to be:
\begin{align}\label{eq:impurity_mode}
    \delta\varphi(t,x)\propto\text{exp}\left(-\epsilon|x|\right)\cos(\Omega t),\quad  \Omega^2=2-\epsilon^2.
\end{align}

To extract the frequency $\tilde\Omega$ and amplitude $A$ of the impurity mode oscillations for the case of the kink-antikink pair interacting with a narrow Gaussian impurity, we use a discrete Fourier transform of $\varphi(0,t)$. We only consider field values at late times ($t>170$) to ensure that the kink and the antikink have entirely left the impurity site. The results are presented in Fig.~\ref{fig:1_dependance}. We find that the dependence of the frequency $\tilde\Omega$ on impurity strength follows the prediction of Eq.~\eqref{eq:impurity_mode}, as shown in Fig.~\ref{fig:1_dependance}c. Here, the maximum observed deviation \red{between $v_{\text{in}}=0.4$ (blue dots in Fig.~\ref{fig:1_dependance}c) and $v_{\text{in}}=0.7$ (red squares in Fig.~\ref{fig:1_dependance}c)} for $\epsilon=0.3$ is $0.35\%$. This suggests that the observed oscillations remaining at the impurity site after a kink-impurity-antikink collision are caused by the excitation of an internal mode native to the impurity. This is further supported by the observation that the frequency of the impurity mode is nearly independent of the initial velocities of the kink and antikink, where we find $\langle \tilde\Omega \rangle \simeq 1.41$ with a maximal deviation of $0.45\%$ for $\epsilon=0.1$ \red{(blue dots in Fig.~\ref{fig:1_dependance}a)} and $\langle \tilde\Omega \rangle \simeq 1.38$  with a maximal deviation of $1.10\%$ for $\epsilon=0.3$ \red{(red squares in Fig.~\ref{fig:1_dependance}a)}.

Even for initial velocities below the critical velocity, we find indications that an impurity mode excitation may remain after a breather decays into the vacuum. Although constraints on the numerical runtime prevent us from observing the complete disappearance of the breather, we do find that for example at $\epsilon=0.5$ and $v_{\text{in}}=0.4$, the field configuration seems to approach a stable oscillation with frequency $ \tilde\Omega \simeq 1.30$. This is in good agreement with the frequency $\Omega \simeq 1.32$ predicted by Eq.~\eqref{eq:impurity_mode} for an impurity mode at this value of the impurity strength.

\section{Repulsive impurity}
In the case of the kink and antikink colliding at the site of a repulsive impurity, with $\epsilon<0$, we find that for initial velocities greater than $v_{\text{cr}}$ the kink and antikink always leave the impurity site and propagate to infinity after a single collision without exciting any impurity mode. For low initial velocities on the other hand, with $v_{\text{in}}<v_{\text{cr}}$, we identify three distinct types of behavior. First, for special values of the initial velocity, we encounter resonances, in which the kink and antikink escape to infinity after a finite number of collisions. This is in stark contrast to kink-impurity interactions~\cite{kivshar,ourownpreprint}, in which resonances are observed only for attractive impurities. Here, we find them at various repulsive impurity strengths and initial velocity, such as $\epsilon=-0.8$ ($v_{\text{in}}=0.771$, $v_{\text{in}}=0.772$) and $\epsilon=-0.4$ ($v_{\text{in}}=0.540$, $v_{\text{in}}=0.555$, $v_{\text{in}}=0.557$). 

\begin{figure*}[tb]
\center{\includegraphics[width=0.8\linewidth]{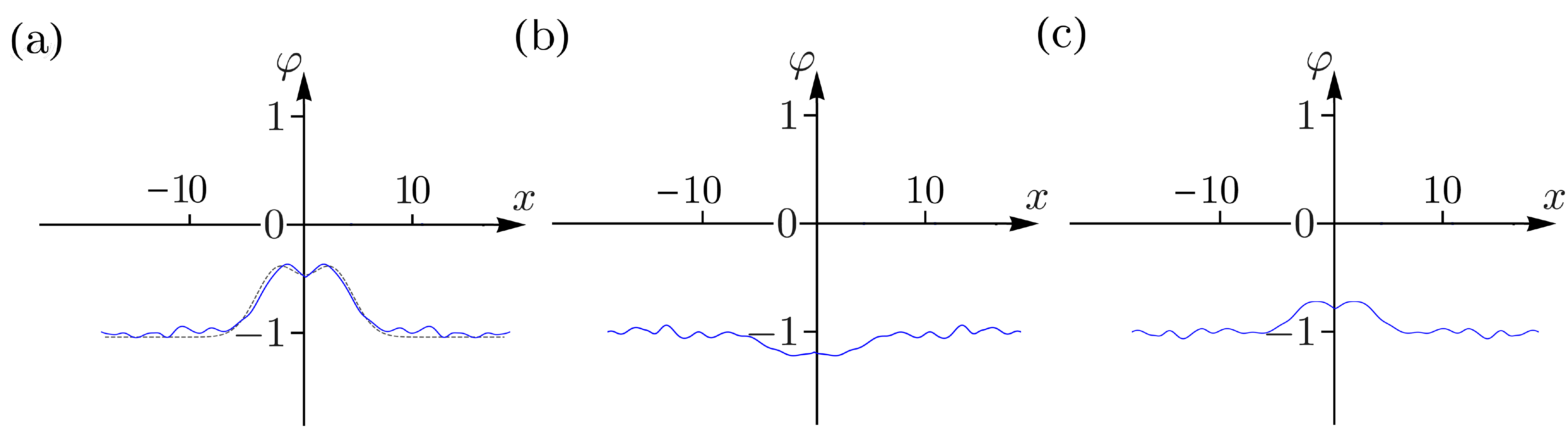}}
\caption{Oscillon-oscillon bound state observed for $\epsilon=-0.2$ and $v_{\text{in}}=0.1$. The dashed line in panel (a) is a fit of the profile to the Ansatz of Eq.~\eqref{eq:oscillon}. Panels (b) and (c) indicate how the profile evolves over time. \red{All panels show the quality behavior at a representative moment after the kink-impurity-antikink collision.}}
\label{fig:oscillon_fig_type1}
\end{figure*}

For other, generic, values of the initial velocity, we observe the formation of so-called oscillon pairs at the impurity site \red{for time much later than kink-impurity-antikink collision}. These may either separate and move away from $x=0$, as shown in Fig.~\ref{fig:oscillon_fig_type2}, or remain close to the impurity site in an oscillon-oscillon bound state (see Figs.~\ref{fig:oscillon_fig_type1} and~\ref{fig:vcr_epsilon_transition}c). An oscillon $\varphi_O(x)$ is a quasi-long-lived bound state of the $\varphi^4$ model that periodically oscillates around one of the vacua~\cite{oscillon3}. At the extreme points in its oscillation, the oscillon has a Gaussian profile:
\begin{align}\label{eq:oscillon}
    \varphi_O(x)=A_O\,\text{exp}\left[-(x-B)^2/C^2\right].
\end{align}
Here, $A_O$, $B$, and $C$ are the amplitude, center, and width of the oscillon respectively. \red{The value of $B$ is only a profile shift in the $x$ direction and does not affect the oscillon shape. Thus, we avoid further discussion of the values of $B$ and focus only on the values of $A_O$ and $C$.} This function gives an excellent fit to the numerically obtained field configurations $\varphi(t,x)$ after kink-impurity-antikink collisions, as shown in Figs.~\ref{fig:oscillon_fig_type2} and~\ref{fig:oscillon_fig_type1}. For $\epsilon=-0.2$ and $v_{\text{in}}=0.30$, the best fit is obtained with $A_O=0.58$ and $C=3.07$, while the case $\epsilon=-0.4$ and $v_{\text{in}}=0.43$ results in $A_O=0.51$ and $C=3.02$. Both values for the width of the oscillon fall within the range $2.90\le C \le 3.54$ that has been shown to yield the longest-lived oscillons in $(3+1)$ dimensions~\cite{oscillon3}. While the frequency of oscillations is close to $\omega_{O}\simeq 1.26$ for both cases considered, the final velocities of the separating oscillons differ significantly ($v_{\text{f}}=0.09$ and $v_{\text{f}}=0.16$, respectively). The impurity strength and initial velocity thus mostly influence the amplitudes and final velocities of the oscillons, while having only a minor influence on their widths or frequencies. As shown in Fig.~\ref{fig:oscillon_fig_type1}, a similarly excellent fit to Eq.~\eqref{eq:oscillon} can be found for the bound state of two oscillons.

\section{The critical velocity}
\begin{figure}[b!]
\center{\includegraphics[width=0.7\linewidth]{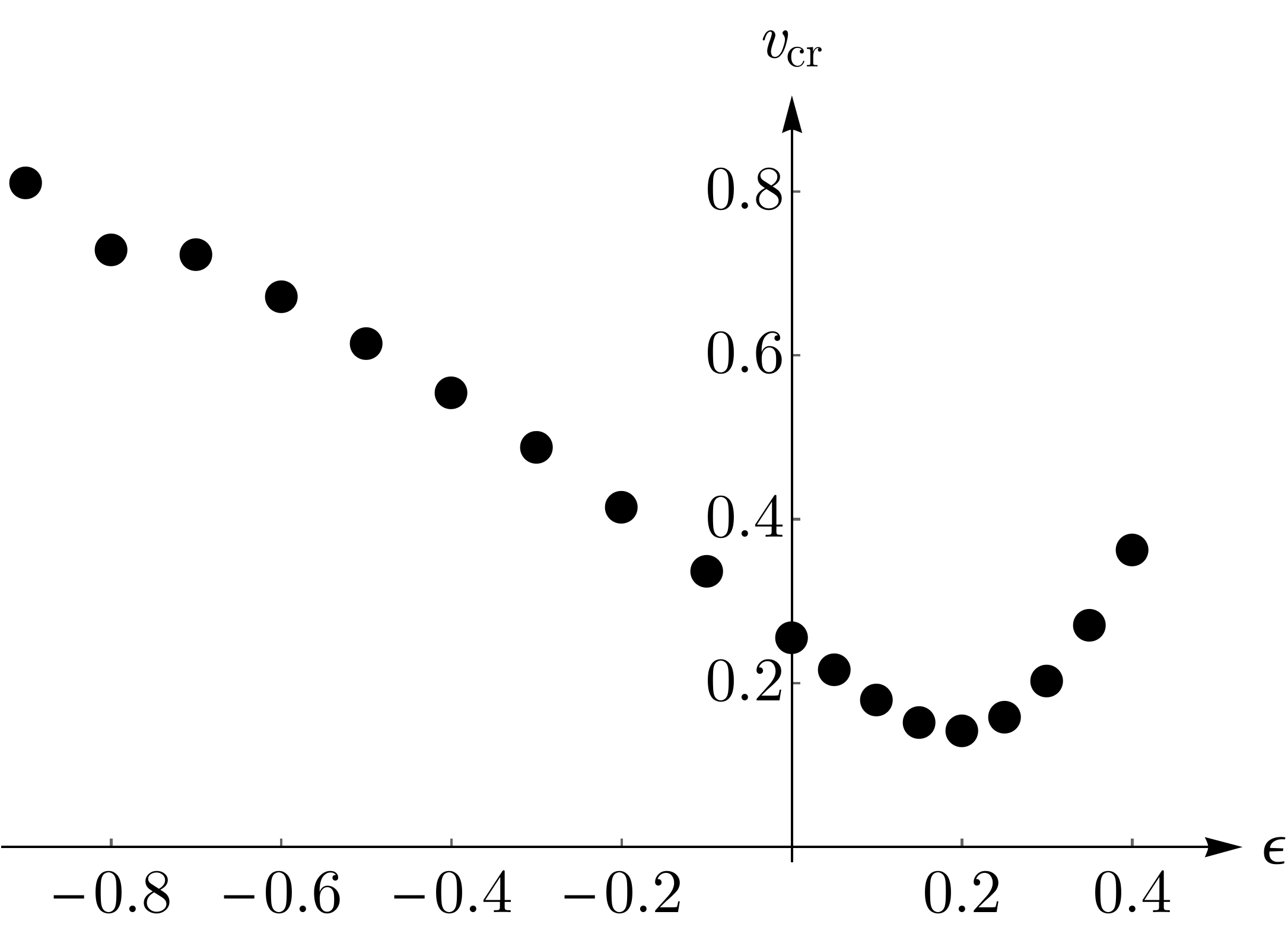}}
\caption{The dependence of the critical velocity $v_{\text{cr}}$ on the impurity strength $\epsilon$.}
\label{fig:vcr_epsilon}
\end{figure}

Figure~\ref{fig:vcr_epsilon} shows the critical velocity as a function of the impurity strength. For repulsive impurities ($\epsilon<0$) colliding kink-antikink pairs with sufficiently low initial velocities form oscillons at the impurity site. Increasing the impurity strength $|\epsilon|$ results in kink-antikink pairs with ever higher initial velocities, forming oscillons. For large positive impurity strength, $\epsilon \gtrsim 0.2$, the kink-antikink pairs are captured and dissipated at the impurity site. Increasing the impurity strength then results in pairs with ever higher initial velocities being captured and dissipated. At low but positive values of $\epsilon$, however, quasi-long-lived breathers are formed, and the critical velocity goes up with decreasing impurity strength. 

The fact that $V_{\text{cr}}$ is a smooth function of $\epsilon$ across the point $\epsilon=0$ \red{($v_{cr}\simeq 0.2598$ \cite{campbell})} and has a minimum at a non-zero positive value of the impurity strength is suggestive of the idea that the field configuration of a bound oscillon pair may be smoothly connected to that of a breather. The region of low, but positive $\epsilon$ then contains quasi-long-lived breathers which obtain part of their stability relative to the bound kink-antikink pairs at high values of the impurity strength from the presence of some oscillon character. This interpretation is consistent with the qualitative observation that oscillon bound states can resonantly decay into propagating kink-antikink pairs within resonance windows for negative $\epsilon$, as well as with the quantitative observation that the oscillon frequency $\omega_O\simeq 1.26$ approximately matches the frequency $\omega_1=\sqrt{3/2}\simeq 1.22$ of the internal vibrational mode for the kink. \red{Similar frequency matching has been observed before in a sinh-deformed $(1+1)$-dimensional $\varphi^4$ model without an impurity~\cite{oscillon8}. There, also, kink-antikink collisions resulted in propagating oscillons rather than bound oscillon-oscillon states for specific values of the initial velocity. In the current kink-impurity-antikink evolution the oscillons appear naturally without deformations, but the matching frequencies between kink and oscillon, and the minimum of $v_{\text{cr}}$ at non-zero $\epsilon$ suggest that it might arise from a similar mechanism to the oscillon-oscillon formation.}

\section{Conclusions}
Studying the collision of a kink and antikink in $(1+1)$-dimensional classical $\varphi^4$ theory in the presence of a single point-like impurity allows us to make several observations about the phenomenology encountered in the $\varphi^4$ model.

First of all, we showed that an internal mode of the impurity could be excited at high initial velocities for an attractive impurity. The observed values for the impurity mode frequency match theoretical and numerical results reported earlier in the study of a single kink interacting with a localized impurity~\cite{kivshar,ourownpreprint}, which shows the excitation to be a property of the impurity that is independent of its excitation conditions.

For small initial velocities of the kink-antikink pair and a repulsive impurity, we observe the emergence of a low-amplitude, quasi-long-lived, oscillating \red{structure}, which we identify to be an oscillon. In contrast to the breather, whose exact structure is still unknown, the oscillon has a Gaussian form at the extreme moments of its oscillations. Originally, oscillons were introduced in $(3+1)$-dimensional $\varphi^4$ theories~\cite{oscillon3}. They were then introduced into simulations in lower dimensions~\cite{oscillon6}, as well as in other models, such as the signum-Gordon model~\cite{oscillon2}, and models of cosmic inflation~\cite{oscillon1}. In all these cases, the oscillons are introduced as an Ansatz for the initial field configuration~\cite{oscillon4,oscillon5}, rather than emerging from physical processes\red{, at least for the $(1+1)$-dimensional $\varphi^4$ model}. Such a natural way for oscillons to appear was recently discovered in the sinh-deformed $(1+1)$-dimensional $\varphi^4$ model~\cite{oscillon8}, after the application of model deformations~\cite{oscillon9,oscillon10}. \red{It also was found within the parametrized Dikand\'e-Kofan\'e DW potential with a parameterized profile that for some values of the parameter reduces to the regular $\varphi^4$ potential~\cite{referee14}.} Here, we show that oscillons also emerge naturally from collisions of kinks and antikinks in the classical $(1+1)$-dimensional $\varphi^4$ model without any deformations. 
Finally, we observe that the critical velocity as a function of impurity strength has a minimum at non-zero positive values of the impurity strength. This is indicative of a smooth transition between bound states of kink-antikink pairs and oscillon-oscillon pairs, with some oscillon character already being present in the breather formed without an impurity present. Such a crossover is consistent with the match between oscillon frequencies and internal kink excitations, which suggests a resonance may arise between the two. 

\acknowledgments
This work is done within the Delta Institute for Theoretical Physics (DITP) consortium, a program of the Netherlands Organization for Scientific Research (NWO) that is funded by the Dutch Ministry of Education, Culture and Science (OCW).

\bibliography{bibliography.bib}

\end{document}